\documentclass[aps,pre,showpacs,showkeys,superscriptaddress,
  floats,floatfix,preprintnumbers, twocolumn,
  longbibliography]{revtex4-1}
\usepackage{amssymb,amsmath}
\usepackage{bm,url,graphics,subfigure,epsfig,rotating,float,tabularx}
\usepackage[normalem]{ulem}
\usepackage{xcolor,soul,enumerate}
\usepackage{hyperref}
\graphicspath{{./}{fig/}}
\def \Tfp   {T_{\rm fp}}
\def \NT   {N^{\rm{T}}}
\def \Ka   {K_{\rm a}}

\def \ri   {\rm{i}}
\def \rj   {\rm{j}}
\def \rs   {\rm{s}}
\def \XX   {\bm{X}}
\def \Ep   {E_{\rm pin}}
\def \mus  {\mu_{\rm s}}
\def \ab   {\alpha\beta}
\def \mus     {\mu_{\rm s}}
\def \Deff     {D_{\rm eff}}
\def \Phic    {\Phi_{\rm c}}
\def \DEinf    {\Delta E_{\infty}}
\def \DE     {\Delta E}
\def \Fs      {F_{\rm s}}

\def \kB   {k_{\rm B}}

\def \ua     {u_{\alpha}}
\def \ub     {u_{\beta}}
\def \sab    {s_{\alpha\beta}}

\def \dela   {\partial_{\alpha}}
\def \delb   {\partial_{\beta}}

\def \qq     {\bm{q}}

\def \mQ {\mathcal{Q}}

\def \uu  {{\bm u}}

\def \kk  {{\bm k}}

\def  \xx  {{\bm x}}
\def  \XX  {{\bm X}}

\def  \ua  {u_{\alpha}}

\def  \ub  {u_{\beta}}

\newcommand{\avg}[1]{\left\langle #1\right\rangle}

\def \mH {\mathcal{H}}

\newcommand{\Eq}[1]{Eq.~(\ref{#1})}

\newcommand{\Fig}[1]{Fig.~(\ref{#1})}
\newcommand{\subfig}[2]{Fig.~(\ref{#1}#2)}

\newcommand{\bfig}{\begin{figure}}
\newcommand{\efig}{\end{figure}}
\newcommand{\bc}{\begin{center}}
\newcommand{\ec}{\end{center}}
\newcommand{\bea}{\begin{eqnarray}}
\newcommand{\eea}{\end{eqnarray}}

\def \Es    {E_{\rm S}}
\def \Eb    {E_{\rm B}}

\def \Sum {\sum}
\def \SUM {\sum}

\def \Xij {X_{\rm ij}}

\def \XXi 	{\XX_{\rm i}}
\def \XXj 	{\XX_{\rm j}}
\def  \XXij  {\XX_{\rm ij}}

\def \lzij  {\ell^{\rm 0}_{\rm ij}}

\def \cAib {{\mathcal A}_i}
\def \LL {{\bm L}}

\newcommand{\SMat}[1]{(see Appendix~\ref{#1})}
\begin{document} 
\title{Anomalous diffusion and effective shear modulus in a semi-solid membrane}
\author{Vikash Pandey}
\affiliation{NORDITA, KTH Royal Institute of Technology and
Stockholm University, Hannes Alfv{\'e}ns v\"ag 12, 10691 Stockholm, Sweden}
\author{Dhrubaditya Mitra} 
\affiliation{NORDITA, KTH Royal Institute of Technology and
Stockholm University, Hannes Alfv{\'e}ns v\"ag 12, 10691 Stockholm, Sweden}

\begin{abstract}
  From the perspective of physical properties, the cell membrane is an
exotic two-dimensional material that has a dual nature: it exhibits
characteristics of fluids, i.e., lipid molecules show lateral
diffusion, while also demonstrating properties of solids, evidenced
by a non-zero shear modulus.  We construct a model for such a
\textit{semi-solid membrane}.  Our model is a fluctuating  randomly triangulated mesh with two different kinds of nodes.  
The solid nodes never change their neighbors, while the fluid nodes 
do. As the area fraction occupied by the solid nodes ($\Phi$) is increased the motion of fluid nodes transition from diffusion to localization via
subdiffusion. Next, the solid nodes are pinned to mimic the pinning
of the plasma membrane to the cytoskeleton. For the pinned membrane,
there exists a range of $\Phi$ over which the model has both a
non-zero shear modulus and a non-zero lateral diffusivity. The bending
modulus, measured
through the spectrum of height fluctuations remains unchanged.
\end{abstract}
\maketitle

The building block of all organisms is the cell.
The cell membrane forms a barrier that separates the cytoplasm from the external
environment.
Composition wise, a typical cell membrane is  approximately 73\% lipid molecules
and about 23\% proteins~\cite{ProteinAreaOcDupuy2008}.
The canonical model of cell membrane is the fluid mosaic model, proposed
by \citet{TheFluidMosaiSinger1972}, who conceptualized it as a fluid membrane
with two leafs made of lipids with diverse inclusions of proteins,
carbohydrates, and cholesterols.
The fluid  membrane is modeled by the  Helfrich energy
functional~\cite{helfrich1973elastic, muller2006biological,
  phillips2012physical} which has energy cost to bending and change in
area but has zero shear elastic modulus.
Minimizing this Hamiltonian with the constraint of fixed volume correctly
captures the common biconcave shape of red blood cells
(RBCs)~\citep{deuling1976red, deuling1976curvature, RestingShapeAPozrik2005}.
Such a model can also be used to correctly interpret micro-pipette aspiration
experiments~\citep{phillips2012physical}.
However, to capture all the different shapes a RBC can have (e.g., discocyte,
stomatocyte, and echinocyte) from a single model it is necessary to consider a
composite model, as noted in Ref.~\cite{hw2002stomatocyte},
\begin{quote}
``where the plasma membrane contributes bending rigidity and the protein-based
membrane skeleton contributes stretch and shear elasticity.''
\end{quote}
Furthermore, it is necessary to include shear elasticity in models of
RBC's to capture their shapes in microfluidic flows~\citep{freund2014numerical}
and to interpret the results of mechanical deformation of RBCs in certain
optical tweezer experiments~\citep{henon1999new,dao2003mechanics,
  MechanicalRespSuresh2006, CellAndMolecuBaoG2003}.
Hence the consensus is that cell membrane of RBCs, and by extension all cells,
have a shear modulus and this emerges due to coupling with the cytoskeleton. 
In contrast, a large body  of experiments~\citep{Ishihara_1987,
  ritchie2003fence,
  james2004compartmentalisation, UltrafineMembrMurase2004,
  ParadigmShiftKusumi2005, manley2008high,Weigel_2011,
  krapf2015mechanisms}
show that the lateral motion of tagged lipid molecules in the cell membrane is
diffusive, although the diffusion may be anomalous.
This is typical feature of a (two-dimensional) fluid. 
Furthermore, the fact that the cell membrane forms tethers provides additional
evidence in support of its fluid nature.
Along similar lines, \citet{shi2018cell} have interpreted the mutual
interaction of two tethers in a cell using a model where the cell
membrane is modeled as a porous
media~\footnote{This is different from modeling \textit{a cell as a porous
media}. } Thus, the cell membrane is an exotic two--dimensional 
material with physical properties akin to  both solids and fluids.
Here we present, for the first time, a minimal model that 
captures this dual nature. 

\begin{figure}
    \centering
        \includegraphics[width=0.95\linewidth]{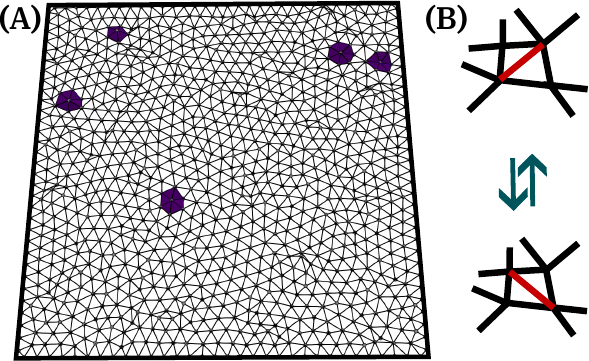}
        \caption{\label{fig:snap0}
          (A) A representative picture of our model: it is a fluctuating,
          randomly triangulated surface where the solid regions are colored in
          purple. It is connected to a frame. (B) An example of bond flips 
          that fluidize our membrane.}
\end{figure}
We start with a model for isotropic and homogeneous
two-dimensional elastic material.
In the Monge gauge the deformation can be described by an in-plane
vector $\uu(x_1,x_2)$ and an out-of-plane displacement $f(x_1,x_2)$.
The Hamiltonian is given by, 
\begin{subequations}
    \begin{align}
      \mH &= \int d\xx \left[{\kappa \over 2}(\nabla^2 f)^2 + \mu s_{\ab}^2  
        + {\lambda \over 2}s^2_{\alpha\alpha} \right]\/,\\
      \text{where}\quad& \sab \equiv \frac{1}{2}
      \left(\dela \ub + \delb \ua + \dela f \delb f \right)\/, 
\end{align}
\label{eq:gen_hamil}
\end{subequations}
is the strain tensor and $\mu$ and $\lambda$ are the two
Lam{\'e} coefficients~\citep{Feynman77,LLelast}.  
A possible Monte Carlo modeling of such a
membrane~\citep{gompper2004triangulated, paulose2012fluctuating,
  auth2019simulating, agrawal2022memc, agrawal2023active} goes as follows.
First, generate a random triangulated grid on a flat surface with $N$ nodes,
see \subfig{fig:snap0}{A}.
Second, update the positions of the points ($\XXi$ at node $i$)
by discretizing the Hamiltonian in \Eq{eq:gen_hamil} over this grid.
The energy has two contribution,
\begin{equation}
  E = \Es + \Eb ,
\label{eq:Etot}
\end{equation}
where the stretching and the bending contributions respectively are,
\begin{subequations}
  \begin{align}
    \Es &= \frac{H}{2}\Sum_{\ri,\rj}\left(\XXij - \lzij\right)^2 \/,
    \label{eq:Es}\\
    \text{where}\quad
    \Xij &\equiv \lvert \XXi - \XXj \rvert \/,\\
\text{and}\quad \Eb &= \frac{\kappa}{2} \SUM_{\ri} \cAib \LL_{\ri} ^2 \/.
  \label{eq:bendE}
  \end{align}
\end{subequations}
The spring constant $H$ sets the two Lam{\'e} coefficients~\cite{seung1988defects,paulose2012fluctuating},
$\cAib$ is the area of the dual Voronoi cell at the
node $\ri$ and $\LL_{\ri}$ is the discretized
Laplacian~\cite{itzykson1986proceedings,
  seung1988defects,  gompper2004triangulated, paulose2012fluctuating,
  agrawal2022memc, agrawal2023active}.
For the rest of this paper we shall call such a membrane \textit{solid membrane}
as it has a finite in-plane shear modulus. 
We use the Metropolis algorithm: we accept a 
move if it results in lowering the energy of the membrane;
a move that increases the energy of the membrane by $\Delta E$
is accepted with the probability $\exp[-\Delta E/\kB T]$ where $\kB$
is the Boltzmann constant and $T$ is the temperature.
We define a single Monte Carlo step as an attempt to move each of the
$N$ nodes at least once on average.
Typically, our production runs use $10^6$ Monte Carlo steps. 

\begin{figure*}
\centering
\includegraphics[width=0.80\linewidth]{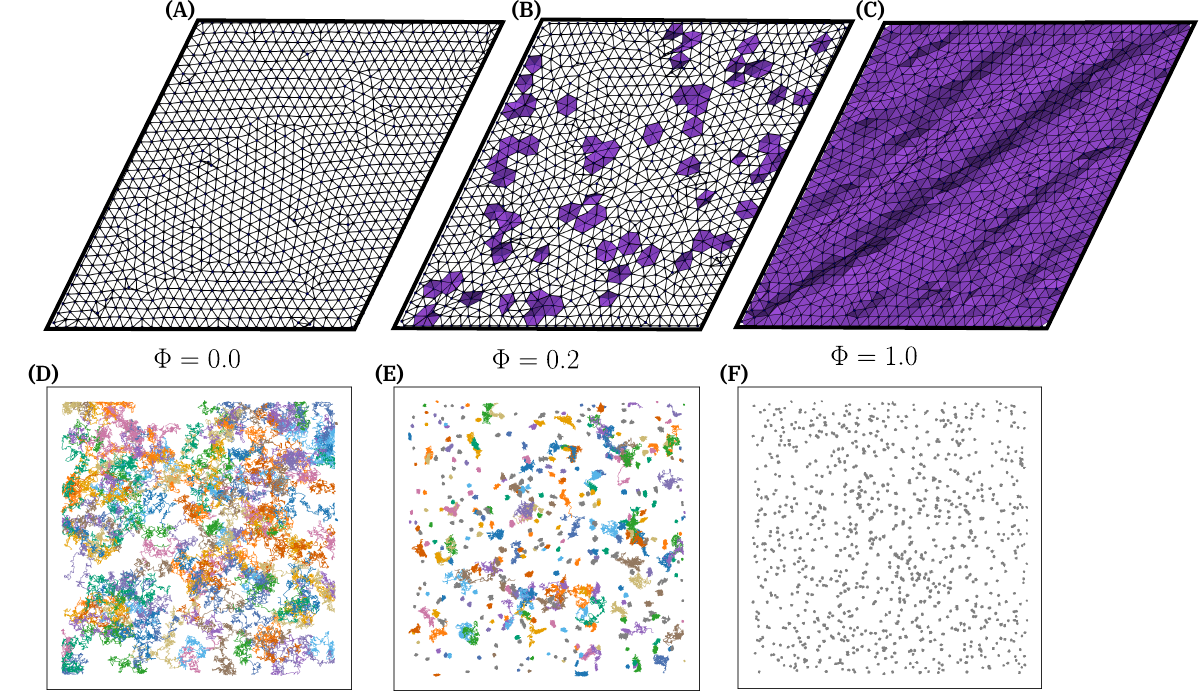}
\caption{\label{fig:snap} (Top panel) Typical snapshots of sheared
  membranes from our simulations.
  The area fraction occupied by solid inclusion is $\Phi$.
  From left to right: (A)$\Phi=0$ (fluid membrane),
  (B)$\Phi = 0.2$ (fluid with solid inclusions), and
  (D) $\Phi=1$ (solid membrane) which shows ridges
  typical of sheared fabric.
  The purple patches in the middle figure represents the solid regions.
  (Bottom panel) Representative trajectories of the nodes for the same
  three values of $\Phi$.}
\end{figure*}

It is possible to \textit{fluidize} this membrane in the following
manner~\citep[see, e.g.,][section 6.2.4]{auth2019simulating}. 
After every $M$ Monte Carlo steps we flip bonds between neighbors --
bond between any two adjacent nodes is broken and a new bond between other
points is formed as shown in \subfig{fig:snap}{B}.
This operation, is always allowed as long as the
minimum number of neighbors of a node  is  greater than four and the length of
new bond is less than a given maximum allowed bond length.

All the parameters used in this study, e.g., the number of nodes,
the total number of Monte Carlo steps, etc, are given in \SMat{sm:sim}.
To estimate the error we use BAGGing (Bootstrap
AGGregation)~\cite{mehta2019high} --  the data ($\Gamma$) are
divided into $n=40$ subsets $\{\Gamma_1, \ldots,\Gamma_n\}$.
For any quantity, e.g., the total energy $E$,
we calculate $n$ mean values, where the $k$-th value is calculated
by averaging over the data in $\Gamma_{k}$.
The standard deviation calculated over all the sets is our error estimate.

The first novelty of our model is to randomly assign a fraction of total nodes
to be \textit{solid} and the rest of them to be fluid.
By this we mean that the bonds of the solid nodes with their neighbors is never
changed while the bonds between two fluid nodes can be exchanged.
Changing the fraction $\Phi$ from $0$ to $1$ we go from a fluid membrane,
to fluid membranes with solid inclusions and finally solid membrane.
In  \subfig{fig:snap0}{A} the triangles formed by the solid points with 
their neighbors are colored.
The fractions of area of the membrane that can be considered solid 
can be defined as $\Phi \equiv \NT_{\rm{s}}/\NT$, 
where $\NT_{\rm s}$ is the number
of triangles associated with the solid points and $\NT$ is 
the total number of triangles.
We study the physical properties of membrane as the $\Phi$ is changed. 

We attach  the nodes lying on the boundary of the membrane to a frame. 
The particles on the frame do not undergo any Monte Carlo move. 
By straining the frame we can impose an external strain on the membrane
and calculate its response. Note that we do not include the triangles
formed with the points on the frame in $\NT$. 

In \subfig{fig:snap}{A} to (C) we show typical snapshots of strained
membrane from our simulations. The panel (A) is for the fluid membrane
($\Phi = 0$) and the panel (C) is for the solid membrane($\Phi = 1$).
The panel (B) is for an intermediate value, $\Phi = 0.2$.  The solid
membrane shows a typical ridges that appear in sheared solid
membranes, e.g., fabrics.  In \subfig{fig:snap}{A} and (B) we allow
bond exchange -- for all nodes in (A) but for $80\%$ nodes in (B).
In \subfig{fig:snap}{D}
to {E} we plot the trajectories of a (fraction of) nodes.  In
\subfig{fig:snap}{D} all the nodes are fluid.  The nodes move by a
large amount.  From (D) to (E) the nodes get more and more localized.
In (E), where no bond exchange is allowed, the nodes are completely
localized.

\begin{figure}
    \begin{center}
        \includegraphics[width=0.95\columnwidth]{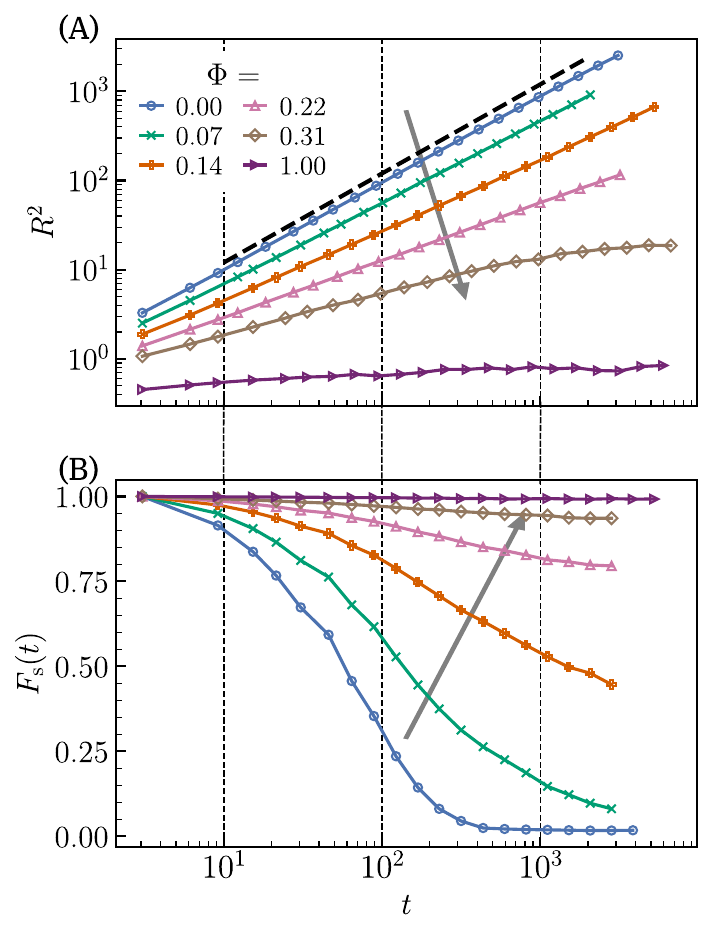}
    \end{center}
    \caption{\label{fig:msd} (A) Log-log plot of the mean-squared-displacement
      $\avg{R^2}$ as a function of $t$ measured in the units of $\tau$.  
      (B) The function $\Fs$ as a function of time $t$.
      The arrow points to the direction of the increase in $\Phi$.}
\end{figure}
In \subfig{fig:msd}{A} we plot the mean square displacement
as a function of time ($t$), 
\begin{equation}
    \avg{R^2(t)} \equiv \avg{[\XX(t) - \XX(0)]^2},
    \label{eq:msd}
\end{equation}
where the symbol $\avg{\cdot}$ shows average over all the nodes. 
The fluid membrane, which corresponds to the top line, shows
$\avg{R^2} \sim t$, i.e., the nodes show diffusion.
As the area fraction $\Phi$ increases we move from 
diffusion to subdiffusion ($\avg{R^2} \sim t^{\beta}$ with $\beta < 1$)
to the case of solid membrane where $\avg{R^2}$ is constant at late 
times -- a typical signature of localization. 
We estimate a $\Phi$ dependent diffusion constant by calculating
\begin{equation}
D(\Phi) \equiv \lim_{t\to\infty}\frac{\avg{R^2}}{t} 
\label{eq:Dphi}
\end{equation}
Clearly there exists a critical value of $\Phi$, $\Phic$
such that for $\Phi > \Phic$ the diffusion
coefficient $D(\Phi) \approx 0$. 

Henceforth, we will use the mean bond length for the initial grid, $b \equiv
\avg{\lzij}$ as our unit of length.  We use the number of Monte Carlo
steps as time and use $\tau \equiv b^2/D (\Phi = 0)$ as the unit of
time.  Energy is measured in the unit of $\kB T$.  In these units the
diffusion constant for the fluid membrane $D(\Phi=0) = 1$.

Measurement of lateral diffusion constant of lipids on giant
unilamellar vesicles (GUVs) which are reinforced with a network of
polymeric filaments (e.g., actin filaments)
shows~\cite{andrews2008actin, heinemann2013lateral, hartel2019giant}
that the diffusion constant decreases as the coupling with the polymer network
increases. Our model reproduces this result qualitatively.
\citet{heinemann2013lateral} models this phenomena with random walkers
 confined within cages created by random network in two dimensions. 
The walkers are allowed to jump across cage barriers with a certain probability.
In contrast, in our model, it is the fluid nodes of the network that
does the random walk while the solid patches acts as impenetrable
barriers.

We also calculate the cumulative probability distribution
function (CDF) of first--passage time for the fluid nodes.
The CDF has exponential tails (\SMat{sm:cdf}) which implies that the
probability distribution functions (PDF) of first--passage times also have
exponential tails. This suggests that the anomalous diffusion we observe is
not a continuous time random walk (CTRW) in which case
the PDF of first--passage times show power--law tails~\cite{metzler2000random,Condamin_2008}.

The anomalous diffusion we observe is akin to diffusion in crowded
environment in two dimensions~\citep{AnomalousTransHoflin2013,
krapf2015mechanisms}, where depending on the fraction of area covered
by the obstacles, the motion of Brownian particles  go from diffusion to subdiffusion for intermediate
times~\citep{bauer2010localization,AnomalousTransHoflin2013}.
 At very late times the
subdiffusive trajectories either becomes diffusive or gets localized
($D \to 0$)~\citep{saxton_1994,bauer2010localization}.
As our
membrane is bounded by a frame we cannot investigate whether the
subdiffusion that we observe goes over to diffusion at 
late times or
not.  \citet{saxton_1994} and \citet{Yethiraj_lateral} have previously
modeled the anomalous diffusion of tagged molecules in 
membranes in lattice and continuum models respectively.
These models have certain qualitative similarities to our
model but there are crucial quantitative differences~\SMat{sm:diff}.

In \subfig{fig:msd}{B} we plot the self-intermediate scattering
function \cite{Van_Hove_1954,Bi_2016}:
\begin{equation}
  \Fs(t) \equiv \avg{\exp\left[i \kk\cdot\XX(t) \right]}
\end{equation}
as a function of time for $\mid\kk\mid = \pi/(8b)$.  The
average is performed over all the angles of of the wavevector $\kk$
and over all the nodes.
For a fluid $\Fs(t) \to 0$ exponentially fast at late times; for a
solid $\Fs(t)$ goes to a non-zero constant at late times.  Again, we
find a transition from fluid to a solid  at a critical
value $\Phi = \Phic$. 

\begin{figure}
    \begin{center}
        \includegraphics[width=0.85\columnwidth]{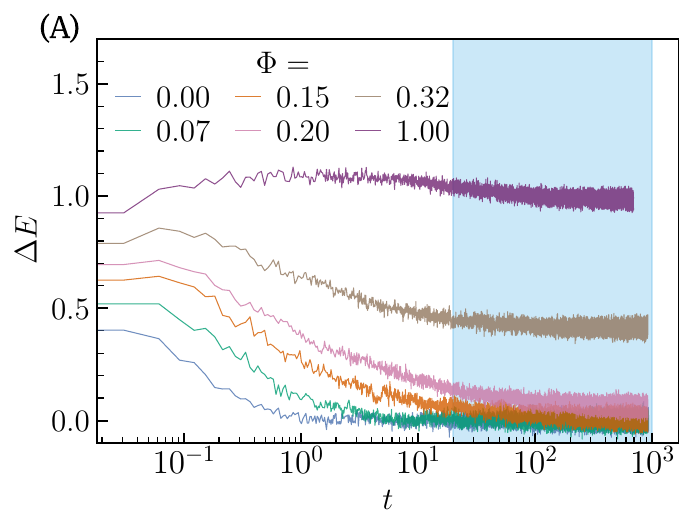}
        \includegraphics[width=0.85\columnwidth]{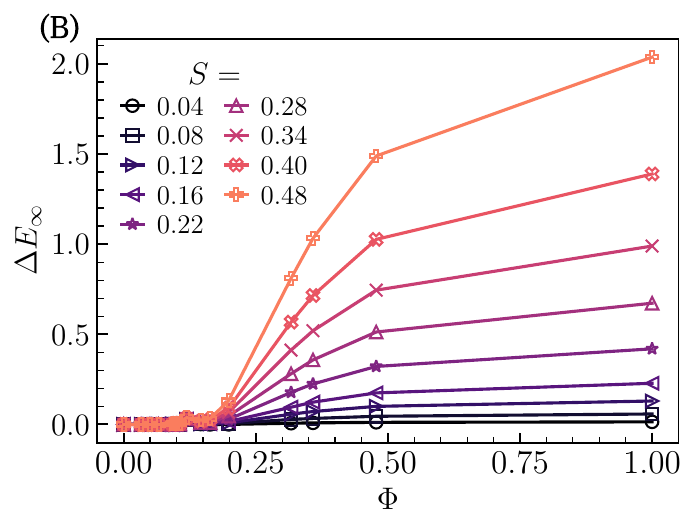}
        \includegraphics[width=0.85\columnwidth]{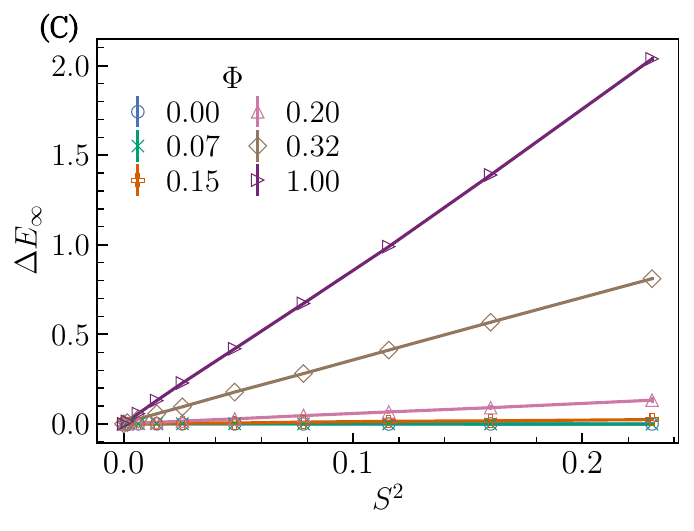}
    \end{center}
    \caption{\label{fig:energy} (A) The change in energy, $\DE$, of the sheared
      membrane as a function of non-dimensional time, for different values of
      $\Phi$ and constant shear strain $S = 0.34$.
      The area fraction $\Phi$ increases from bottom to top.
      The mean energy at late times is $\DEinf$, (calculated as an average
      over the shaded region ) which is a function of both
      $S$ and $\Phi$.
      (B) The change in energy at late times, $\DEinf$, as a function of $\Phi$
      for different values of $S$.
      For $\Phi$ smaller than certain critical value $\Phic \approx 0.23$,
      $\DEinf = 0$ for all values of $S$.
      (C) The change in energy at late times, $\DEinf$, as a function of $S^2$
      for different values of $\Phi$. For $\Phi < \Phic$, $\DEinf = 0$.
      The slopes of these lines are the effective shear modulus $\mu(\Phi)$.
      The errorbars in (B) and (C) are smaller than the
      size of the symbols.}
\end{figure}

Next we study the response of the membrane to imposed linear shear
strain.  We deform the frame such that a point at location $x,y$ on
the frame is displaced by $Sy,0$. The strain of the frame is given by
$S$.  In \subfig{fig:energy}{A} we plot the change in energy of the
membrane $\DE$ as a function of time for different values of $\Phi$
and for a single strain.  The energy is maximum at $t=0$.  As time
progresses the energy relaxes and at late times reaches a constant
value $\DEinf$ with fluctuations.  For $\Phi < \Phic$ we find that the
membrane relaxes such that at late time $\DEinf = 0$.  In other words,
there is no energy cost to shear strain.  These are examples of a (two
dimensional) fluid membrane.  But for $\Phi > \Phic$ at late times the
$\DEinf$ reaches a constant non-zero value (with thermal fluctuations).
This is a feature of solid membranes which can sustain shear strain.
In \subfig{fig:energy}{B} we plot the change in energy at late times,
$\DEinf$, as a function of $\Phi$ for different values of strain, $S$.
For $\Phi < \Phic$, $\DEinf$ is zero for all strains.  For $\Phi >
\Phic$, $\DEinf$ is finite and also increases with $S$.
In \subfig{fig:energy}{C} we plot $\DEinf$ as a function of $S^2$ for
different values of $\Phi$.  We obtain $\DEinf \propto S^2$, the
constant of proportionality is the effective shear modulus $\mu$ which
depends on $\Phi$.

\begin{figure}
    \begin{center}
        \includegraphics[width=0.90\linewidth]{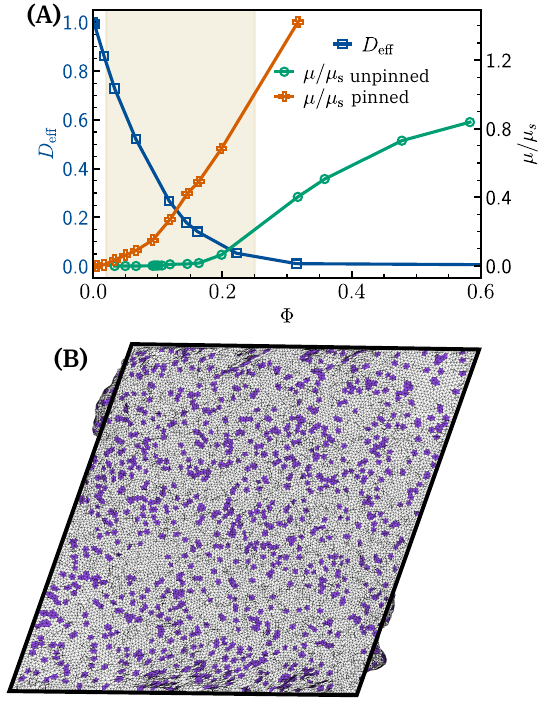}
    \end{center}
    \caption{\label{fig:Dmu}
      (A) The diffusion coefficient $D$, (left axis) and the effective
shear modulus $\mu$ (right axis) as a function of area fraction
$\Phi$.  The coefficient $D$ is normalized by the diffusivity of the
fluid membrane, $D(0)$.  The shear modulus $\mu$ is normalized by the
shear modulus of the solid membrane $\mus$.  The diffusion coefficient
is zero for $\Phi > \Phic$, with $\Phic \approx 0.23$.  The shear
modulus is zero for $\Phi < \Phic$.  Both of them show a rigidity
transition at $\Phi = \Phic$.  The diffusivity of the pinned membrane
(not shown here) is equal (within error bars) to the non-pinned one.
The orange curve shows $\mu/\mus$ for the pinned membrane, for3 which
$\mu >0$ for $\Phi \gtrsim 0.04$.  Thus for $0.04 <\Phi < \Phic$, the
shaded region, the pinned membrane has physical properties akin to  
both solids and fluids.
The error bars in the data are smaller than the size of
the symbols.  (B) A typical snapshot of the strained pinned membrane
for $\Phi = 0.2$ and $S = 0.34$.}
\end{figure}

In \subfig{fig:Dmu}{A}, we plot both the diffusivity obtained from
\subfig{fig:msd}{A} and the shear modulus $\mu$ obtained from
\subfig{fig:energy}{C} as a function of the
area fraction $\Phic$.
We measure $D$ in units of $D(\Phi=0)$ and measure $\mu$ in units of 
$\mus \equiv \mu(\Phi=1)$ 
-- the shear modulus of the solid membrane. 
As $\Phi$ increases $D$ decreases with $D\approx 0$ for
$\Phi >\Phic$ with $\Phic \approx 0.22$.
Simultaneously $\mu \approx 0$ for $\Phi < \Phic$. 
The shear modulus $\mu$ increases as $\Phi$ increases beyond $\Phic$. 
There is a very tiny range of $\Phi$ around $\Phic$ for which both $\mu$
and $D$ are non-zero although both are much smaller than unity.
We conclude that there is a transition of the membrane
from fluid behavior to solid at $\Phi = \Phic$,
characterized by the measurement of diffusion constant and the shear modulus. 
This is a typical rigidity transition.
The membrane is either a solid ($\Phi > \Phic$) or a fluid ($\Phi < \Phic$)
-- it does not have the dual nature -- both a solid and a fluid --   exhibited by cell membranes. 

We now introduce the second novelty in our model. 
For all the cases where $\Phi$ is not equal to zero or unity, i.e.,
for the cases where there are both solid and fluid nodes in the model,
we introduce a harmonic pinning potential that pins (anchors) only the
solid points to the $x_1$--$x_2$ plane. 
The additional term in the energy of the membrane is 
\begin{equation}
    \Ep = \Sum_{\ri = \rs}\frac{K}{2}\left(\XX_{\ri} - \XX_{\ri}^0\right)^2\/,
    \label{eq:harmonic}
\end{equation}
where the suffix `$\rs$' denotes the solid points.
The vector $\XX_{\ri}^0$ is the initial position of the $\ri$-th node
and $K$ the spring constant of the pinning potential. 
The goal is to mimic the attachment of the membrane to the cytoskeleton. 
In \subfig{fig:Dmu}{B} we show a typical snapshot of the sheared 
\textit{pinned} membrane. 
We find that the pinning has no effect on $\avg{R^2}$, it is indistinguishable
from $\avg{R^2}$ of the unpinned membrane~\SMat{sm:diff}. 
Consequently the lateral diffusion coefficient depends on $\Phi$
alone -- it is independent of whether the solid points are pinned or not. 
But, pinning significantly changes the shear modulus $\mu$.
To \subfig{fig:Dmu}{A} we now add the plot of $\mu$ [normalized by 
$\mu(\Phi=1)$] for the pinned membrane as a function of $\Phi$.
For $\Phi > 0.04$ the the pinned membrane has a non-zero $\mu$.
The shear modulus of the pinned membrane is always higher than the unpinned
one. 
Remarkably there is a large range of values of $\Phi$, $0.04 < \Phi < 0.2$
over which the membrane is a fluid, characterized by the diffusivity of its
nodes, but also solid if the shear elastic modulus $\mu$ is measured.
This range of $\Phi$ is shaded in \subfig{fig:Dmu}{A}.
We have now reached our goal, a minimal model of the cell membrane
that can capture both its solid and fluid aspects. 

How does pinning affect bending and area modulus of the membrane?
The bending modulus is not affected by pinning.
To show this, we calculate the power--spectra of height
fluctuations:
\begin{equation}
  G(q) = \avg{f(\qq)f(-\qq)}\/.
  \label{eq:S}
\end{equation}
At large $q$ we obtain $G(q) \sim \kappa q^4$ where the bending modulus
$\kappa$ is the same for all area fraction $\Phi$,~\SMat{sm:kappa}.
Note that pinning does change $G(q)$ at small $q$: for the solid
and the fluid membrane, none of which are pinned,
$G(q) \sim q^2$ at small $q$ whereas for the pinned membranes
$G(q)$ becomes almost independent of $q$ as $\Phi$ increases. 

The area modulus can be written as a function of the two Lam{\'e}
coefficients: $\Ka = 2(\mu + \lambda)$, of which we have already
calculated $\mu$.  To calculate $\lambda$ we deform our frame such
that its shape remains the same (a square) but the area changes.  The
change in energy of the membrane at late times gives $\lambda$.  As
expected, we find that both the (unpinned) solid and fluid membrane
shows the same response to stretching; pinning increases
$\Ka$~\SMat{sm:Ka}.

In summary, in this paper we introduce a model of membrane that has
physical properties akin to  both solids and fluids.
In particular, within a certain parameter range, the fluid nodes of the
membrane show (anomalous) diffusion while the membrane as a whole have a
non-zero elastic shear modulus.
The key contribution to the elastic shear modulus comes from the pinning
of the solid nodes to a flat surface.
We emphasize that although motivation behind our model is the cell membrane,
 we do not attempt to capture all the incredible complexity of natural biological
membranes  in this model.
A simpler experimental analog of our model is a giant unilamellar vesicle (GUV)
with cytoskeleton~\citep{murrell2011spreading, tsai2011encapsulation,
  vogel2013design, hartel2019giant, liu2009biology} -- a step towards
building synthetic cells. 
In such composite systems, also in cells, the coupling of the cytoskeleton to
the membrane can be tuned.
Experiments~\citep{tsai2011encapsulation} show that inclusion of cytoskeleton
has minimal effect on the bending modulus (measured by the
fluctuation spectrum of height field).
We find the same negative result.
A different technique (hydrodynamic tube--pulling) shows that the effective
cortical tension increases~\citep{guevorkian2015mechanics}.
The tension is proportional to the area modulus $\Ka$ which
in our model does increase. 
To the best of our knowledge, the effective shear modulus of such systems have
not been measured yet.
Note that, while comparing our model with these experiments we expect at best
a qualitative agreement because we do not take into account either the active
nature of the cytoskeleton or its viscoelastic property.
It is straightforward to include activity and study
how it influences the mechanical properties in our
model along the lines described in Refs.~\citep{agrawal2023active}.

From the point of view of soft matter physics, our model describes an
exotic two dimensional material with physical properties akin to  both solids and fluids.

\acknowledgements
We thank  Vipin Agrawal, Bidisha Sinha and Ananyo Maitra for discussions. 
The figures in this paper are plotted using the free software
matplotlib~\citep{Hunter:2007}.  
We use FFTW~\cite{FFTW05} for fast Fourier transform.
DM acknowledges the support of the Swedish Research Council Grant No.
638-2013-9243 and 2016-05225.
The simulations were enabled by resources provided by the Swedish National
Infrastructure for  Computing at PDC partially funded by the Swedish Research
Council through grant agreement  no. 2018-05973.
%

\clearpage
\newpage
\appendix
\section{Details of simulation}
\label{sm:sim}
\begin{table}[!ht]
\begin{tabular}{|c|c|}
  \hline
	$N$  & 65536 (diffusion expt.)  \\ 
	& 16384 (shear expt.)    \\ 
\hline
$H$  & $10^{4}$  \\ 
\hline
$\kappa$  & 4.0  \\ 
\hline
$\Phi$  & 0.0, 0.002, 0.017, 0.034, 0.051, \\
        & 0.068, 0.094, 0.12, 0.14, 0.17, \\
        & 0.21, 0.31, 0.35, 0.48 \\
\hline
$S$  & 0.04, 0.08, 0.12, 0.16, 0.22,\\ 
     &  0.28, 0.34, 0.40, 0.48 \\
\hline
  $K$   & 200 \\
\hline
\end{tabular}
\caption{\label{sm:tabparam}The parameters used in our simulations:
  $N$ is the total number of nodes;
  $H$ is spring constant of the harmonic spring that connects to
  neighbouring nodes;
  $\kappa$ is the bending modulus;
  $\Phi$ is the area fraction occupied by the solid nodes;
  $S$ is the externally imposed strain on the frame;
  $K$ is the spring constant of the harmonic pinning potential.
  Our production runs are  typically for $10^6$ Monte Carlo steps.
The length of a side of the square frame is $\sqrt{N}b$
where $b$, the average bond length, is our unit of length.
The parameters, $H$, $\kappa$ and $K$ are measured in the
unit of $\kB T$; $N$, $\Phi$ and $S$ are dimensionless. 
}
\end{table}
\section{Diffusion}
\label{sm:diff}
The anomalous diffusion observed in a heterogeneous medium like plasma
membrane is often attributed to crowding due to
macromolecules~\cite{AnomalousTransHoflin2013,krapf2015mechanisms}.
The effect of crowding is modeled by confining a random-walker in a space
filled with immobile obstacles.
This model has been studied in a
lattice \cite{saxton_1994}, where certain sites are inaccessible to a
walker or in a continuous medium, where a Brownian walker diffuses
through the gaps between the obstacles-- also known as Lorentz
model~\cite{Yethiraj_lateral,bauer2010localization}.
As the concentration of the obstacles is increased the walker
goes from diffusion to subdiffusion for intermediate times.
At very late times the
subdiffusive trajectories either becomes diffusive or gets localized
($\Deff\to 0$)~\citep{saxton_1994,bauer2010localization}.

\begin{figure*}
    \begin{center}
        \includegraphics[width=0.38\linewidth]{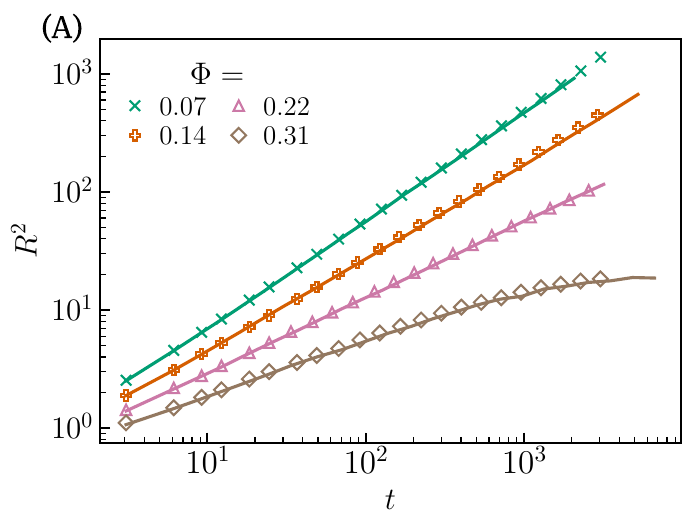}
        \includegraphics[width=0.38\linewidth]{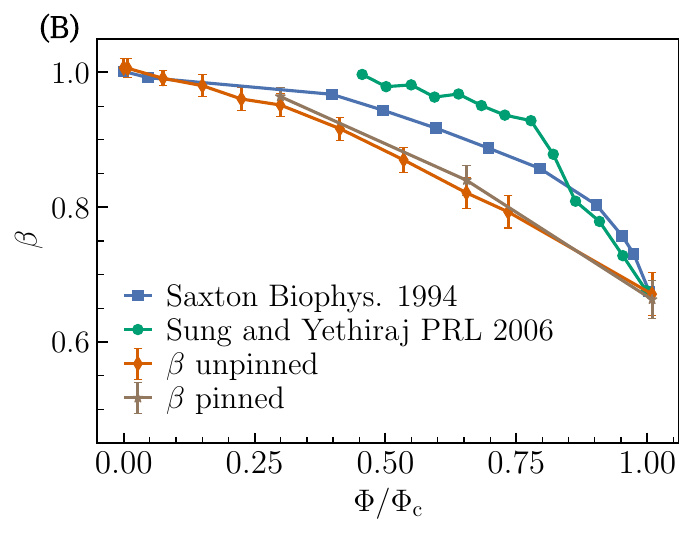}
        \includegraphics[width=0.38\linewidth]{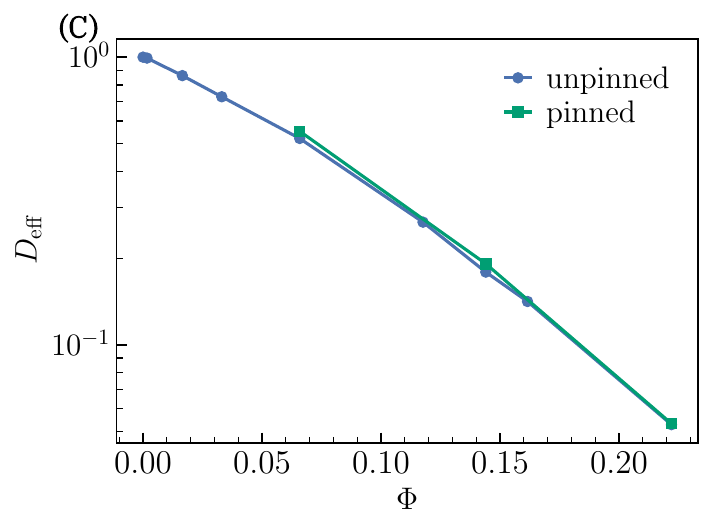}
        \includegraphics[width=0.38\linewidth]{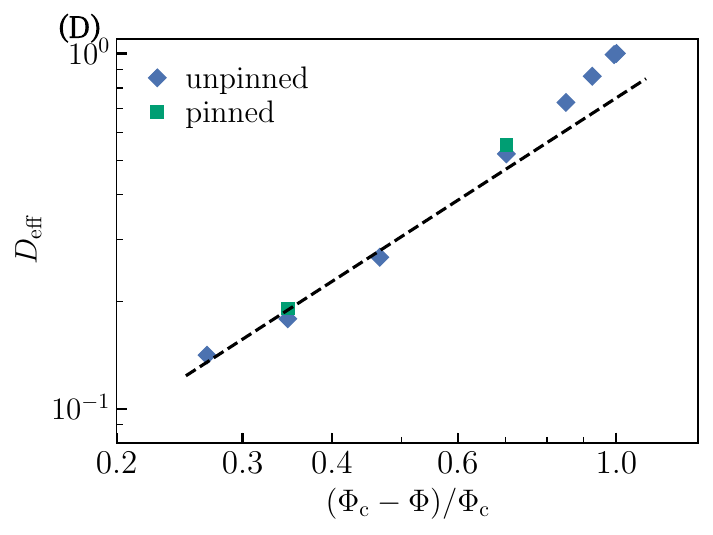}
    \end{center}
    \caption{\label{fig:Msd} (A) Comparison of mean square displacement, of the fluid point,
    between the two cases: (A) solid nodes are pinned (symbols) and (B) they are unpinned (lines). 
    (B) The exponent $\beta$ as obtained from mean square displacement. For comparison we overlay 
    the data obtained by \citet{Yethiraj_lateral} and \citet{saxton_1994}. 
      (C) Semi-log plot of $\Deff$ versus $\Phi$. (D)
      Log-log plot of $\Deff$ versus
	$(\Phic-\Phi)/\Phic $. The dashed line, with an exponent $1.33$
      is the corresponding scaling relation for the
      Lorentz model. }
\end{figure*}

There are qualitative similarities between these models and ours.
In particular, the fluid nodes corresponds to the Brownian walker
while the solid patches act as obstacles.
We run two kinds of simulations:
\begin{enumerate}[(A)]
\item The solid nodes and the patches around them move around.
\item The solid nodes are pinned.
\end{enumerate}
We call the former unpinned (or unanchored) and the latter pinned.
It turns out that the diffusive behavior of the fluid nodes
is independent of whether the solid nodes are pinned or not.
In \subfig{fig:Msd}{A} we show the mean-square-displacement ($\avg{R^{2}}$)
of the fluid nodes for the case where the solid nodes are pinned (open
symbols) and unpinned (continuous lines).
They fall on top of each other. 

We interpret this data in two ways.
First we consider the transition from diffusion to subdiffusion
as $\Phi$, the area fraction occupied by the solid patches, is
changed. 
We fit the means square displacement (as a function of time) with
\begin{equation}
  \avg{R^2} \sim t^{\beta}
\end{equation}
to extract $\beta$, where $\beta = 1$ corresponds to diffusion.
For $\Phi = 0$ we obtain $\beta = 1$.
Beyond a critical value of $\Phi$, $\Phic$ the trajectories
are localized, i.e., $\avg{R^2}$ goes to a constant at late times.
It is difficult to determine $\Phic$ accurately as the
diffusion is limited to a fixed box.
From our data we estimate $\Phic \approx 0.23$.
In \subfig{fig:Msd}{B} we plot $\beta$ as a function of
$\Phi/\Phic$.
 In the same figure, we also compare our results with
two earlier works.
As the models used in these works cannot be mapped to ours or
to each other we identify our $\Phi$ with  the fraction of inaccessible sites in
Ref.~\cite{saxton_1994}
and the area occupied by immobile obstacles in
Ref.~\cite{Yethiraj_lateral}.
Our results are not too different from these earlier works.
For example, in Ref.~\citep{saxton_1994}, $\Phic$ is
the equal to percolation transition of the
lattice.  \citet{Yethiraj_lateral} reports $\Phic=0.22$.

The second way to interpret the data of $\avg{R^2}$ as a function of time
is to calculate a $\Phi$ dependent diffusivity
\begin{subequations}
  \begin{align}
    D(\Phi) &\equiv \lim_{t\to\infty}\frac{\avg{R^2}}{t}\/,  \\
    \text{and}\quad   \Deff &\equiv \frac{D(\Phi)}{D(\Phi=0)}\/,
  \end{align}
  \label{eq:dphi}
\end{subequations}
the effective diffisivity. 
We plot $\Deff$ as a function of $\Phi$ in \subfig{fig:Msd}{C}.
For the Lorentz model,
\begin{equation}
  \Deff \sim {\left(\frac{\Phic-\Phi}{\Phic}\right)}^{\gamma}
\end{equation}
with  $\gamma \approx 1.33$~\cite{bauer2010localization}.
It is not clear from our data whether we see such a behavior.
We plot $\Deff$ as a function of $\Phi$ in both a log--lin
plot~\subfig{fig:Msd}{C} and a log--log plot~\subfig{fig:Msd}{D}.
The precise dependence of  $\Deff$ near
$\Phi \approx \Phic$ is not central to our work. We
leave it for a future investigation.

\subsection{Waiting time distribution}  
\label{sm:cdf}
The underlying mechanism for the anomalous diffusion can be described by
various theoretical models such as fractional Brownian motion (fBM), and
continuous time random walk (CTRW)~\cite{metzler2000random, Condamin_2008,
  AnomalousTransHoflin2013}.
In CTRW, the time taken by a walker to take a step is not constant
but is drawn from a distribution which has a power-law tail.
The exponent characterizing this waiting time distribution determines
to the exponent of the mean square displacement,
$\beta$~\cite{metzler2000random}.

\begin{figure}
    \begin{center}
        \includegraphics[width=0.82\linewidth]{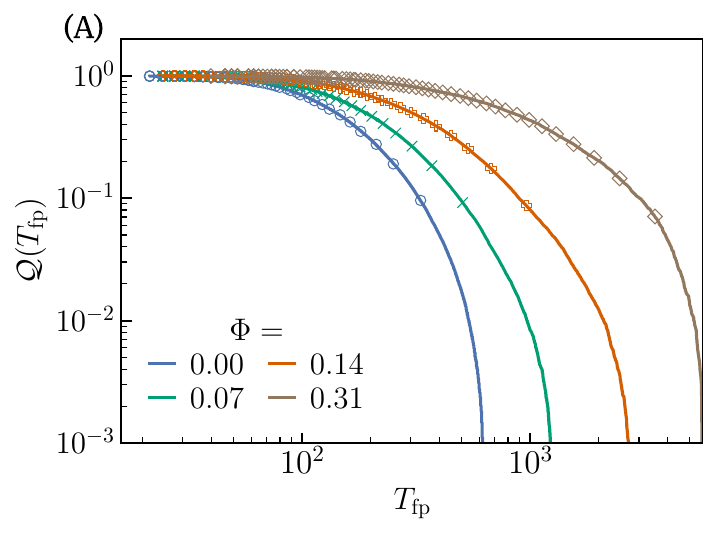}
        \includegraphics[width=0.82\linewidth]{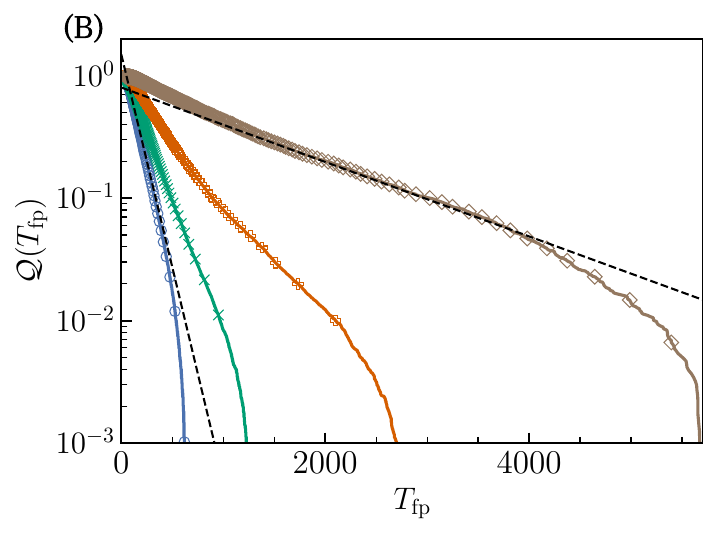}
    \end{center}
    \caption{\label{fig:tfp} (A) Log-log and (B) semi-log plot of the
cumulative distribution function $\mQ(\Tfp)$ of the first-passage-time $\Tfp$, for
different $\Phi$. The dashed lines in panel (B) indicates $\exp(-C
T_{\rm fp})$, where C is chosen to best fit the curves. }
\end{figure}

The exponent of the waiting time distribution can be estimated by studying
the first passage time statistics~\cite{Condamin_2008}.
From the trajectory of fluid nodes
we calculate the cumulative probatility distribution function, $\mQ$,
of the first passage time $\Tfp$, which defined as time taken by a
fluid particle to exit a circle of radius $11b$ \textit{for the first time}.
We plot the distribution in \subfig{fig:tfp}{A} in log--log scale.
It is clear that the tails are not power laws. 
The same behavior was also observed in the model of
Ref~\citep{Yethiraj_lateral}.
The tail of the $\mQ$ can be best approximated by an exponential,
see \subfig{fig:tfp}{B} which implies the probability distribution of
$\Tfp$ also have an exponential tail.
\section{Measurement of elastic constants}
\label{sm:Ka}
We measure the two two Lam{\'e} coefficients $\lambda$ and $\mu$
by deforming all the in-plane coordinates, including the coordinates
for the frame. 
This increases the total energy for the membrane.
We then hold the frame fixed and let the membrane evolve
through Monte Carlo steps.

\begin{figure}
  \centering
    \includegraphics[width=0.95\linewidth]{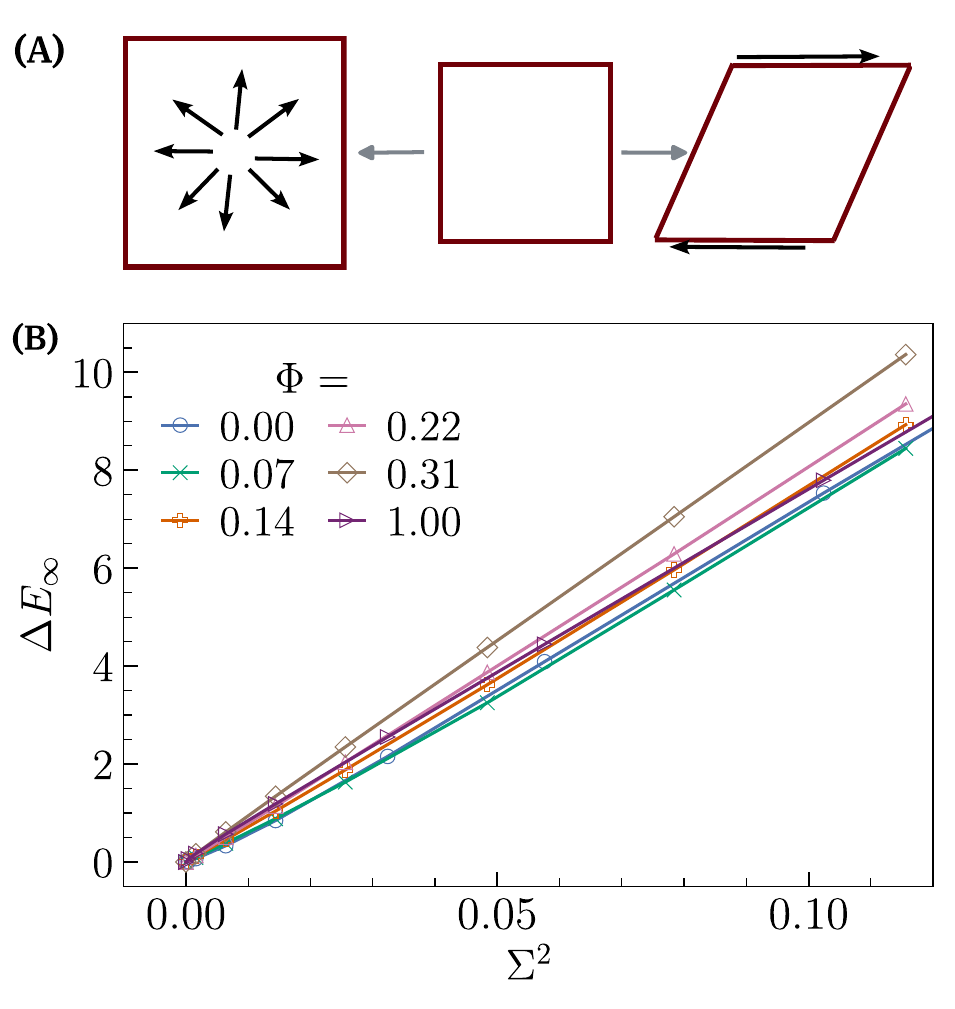}
    \caption{(A) The deformation of the frame:
    Middle figure shows the  undeformed frame, left shows uni-axial
    stretching and right shows shear deformation.
    The frame is held fixed and the points in the membrane are allowed
    to relax via Monte Carlo steps.
    (B) The change in energy relative to the initial configuration
    at late times, $\DEinf$ as a function of the $\Sigma^2$.}
\label{fig:exp}
\end{figure}

To measure $\mu$ we do the following deformation:
\begin{equation}
    \begin{bmatrix} x_1 \\ x_2 \end{bmatrix}	\to 
    \begin{bmatrix} x_1 \\ x_2 \end{bmatrix}	+
    \begin{bmatrix} 0 & {S}\\ 0 & 0 \end{bmatrix}
    \begin{bmatrix} x_1 \\ x_2 \end{bmatrix},  
    \label{eq:strain}
\end{equation}
To measure $\lambda$ we perform the following deformaion. 
\begin{equation}
    \begin{bmatrix} x_1 \\ x_2 \end{bmatrix}	\to 
    \begin{bmatrix} x_1 \\ x_2 \end{bmatrix}	+ 
    \begin{bmatrix} \Sigma & 0\\ 0 & \Sigma \end{bmatrix}
    \begin{bmatrix} x_1 \\ x_2 \end{bmatrix}.
    \label{eq:exp}
\end{equation}
to achieve uni-axial expansion.
The two deformation are shown in \subfig{fig:exp}{A}.

We show the change in energy relative to the initial configuration
at late times, $\DEinf$ versus $\Sigma$ in \subfig{fig:exp}{B}.
We find that both solid and fluid membrane 
show the same response to stretching.
We note that the $\DEinf$ is larger,
at same $\Sigma$, for the cases of pinned membrane
which implies $\Ka$ increases with pinning.

\subsection{Measurement of bending modulus}
\label{sm:kappa}
We remind the reader the theory of elasticity of
an isotropic and homogeneous two--dimensional elastic material.
In the Monge gauge the deformations can be described by an in-plane
vector $\uu(x_1,x_2)$ and an out-of-plane displacement $f(x_1,x_2)$.
The Hamiltonian is given by 
\begin{subequations}
    \begin{align}
      \mH &= \int d\xx \left[{\kappa \over 2}(\nabla^2 f)^2 + \mu s_{\ab}^2  
        + {\lambda \over 2}s^2_{\alpha\alpha} \right]\/,\\
      \text{where}\quad& \sab \equiv \frac{1}{2}
      \left(\dela \ub + \delb \ua + \dela f \delb f \right)\/, 
\end{align}
\label{eq:hamil}
\end{subequations}
is the strain tensor and $\mu$ and $\lambda$ are the two
Lam{\'e} coefficients~\citep{Feynman77,LLelast}.  
\begin{figure}
    \centering
        \includegraphics[width=0.95\linewidth]{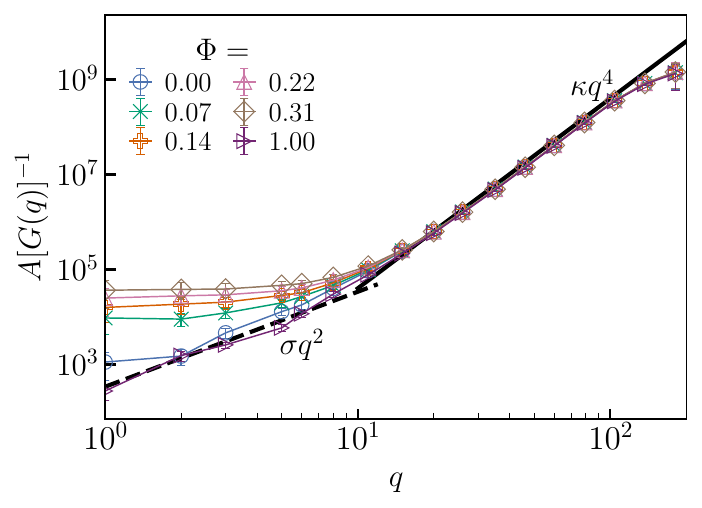}
        \caption{Log-log plot of the power spectra of the height fluctuations.
          The black dashed line shows the $\sigma q^{2}$ and the
           $\kappa q^{4}$ scaling is shown using a continuous line of the same color. }
\label{fig:spec}
\end{figure}

We calculate the spectra for the out of plane fluctuations
$f({\bm x})$ where the in--plane co-ordinates $\xx = (x_{1},x_{2})$.
For small fluctuations, the power spectra for the fluctuations, $G(q)$, 
is given by,
\begin{equation}
     G(q) = { A  \kB T \over \left(\kappa q^4 +  \sigma q^2\right)}, 
    \label{eq:G}
\end{equation}
where $A$ is the area of the projection of the membrane on to the
$x_1$--$x_2$ plane.,
$\kappa$ is the bending modulus
and
$\sigma = ({\mu + \lambda})\left\langle[\cAib/\cAib(t=0) - 1] \right\rangle$
is the effective tension~\citep{phillips2012physical}.
We show the fluctuation spectra from our simulations 
in \Fig{fig:spec}.
For both $\Phi = 0$ (fluid membrane)  and $\Phi = 1$ (solid membrane) we can
identify both the
$\kappa q^{4}$ at large $q$ and  and $\sigma q^{2}$ at small $q$. 
At intermediate $\Phi$, the $q^{2}$ scaling disappears as solid regions are
pinned to the substrate.


\end{document}